\documentclass[aps,pra,superscriptaddress,notitlepage,nofootinbib,longbibliography, colorlinks=true]{revtex4-2}
\usepackage{mathrsfs,graphicx,amsfonts,amssymb,amsmath,mathtools}
\usepackage[dvipsnames]{xcolor}
\usepackage[detect-all=true,group-minimum-digits=4]{siunitx}
\usepackage[colorlinks,allcolors= blue]{hyperref}
\usepackage{cleveref}
\begin{document}

\title{Nonreciprocal entanglement in a molecular optomechanical system}

\author{E. Kongkui Berinyuy}
\email{emale.kongkui@facsciences-uy1.cm}
\affiliation{Department of Physics, Faculty of Science, University of Yaounde I, P.O.Box 812, Yaounde, Cameroon}

\author{Jia-Xin Peng}
\email{JiaXinPeng@ntu.edu.cn}
\affiliation{School of Physics and Technology, Nantong University, Nantong, 226019, People’s Republic of China}

\author{A. Sohail}
\email{amjadsohail@gcuf.edu.pk}
\affiliation{Department of Physics, Government College University, Allama Iqbal Road, Faisalabad 38000, Pakistan}

\author{P. Djorwé}
\email{djorwepp@gmail.com}
\affiliation{Department of Physics, Faculty of Science, University of Ngaoundere, P.O. Box 454, Ngaoundere, Cameroon}
\affiliation{Stellenbosch Institute for Advanced Study (STIAS), Wallenberg Research Centre at Stellenbosch University, Stellenbosch 7600, South Africa}

\author{A.-H. Abdel-Aty}
\email{amabdelaty@ub.edu.sa}
\affiliation{Department of Physics, College of Sciences, University of Bisha, Bisha 61922, Saudi Arabia}
\affiliation{Physics Department, Faculty of Science, Al-Azhar University, Assiut 71524, Egypt}

\author{N. Alessa}
\email{naalessa@pnu.edu.sa}
\affiliation{Department of Mathematical Sciences, College of Science, Princess Nourah bint Abdulrahman University, P.O. Box 84428, Riyadh 11671, Saudi Arabia}

\author{K.S. Nisar}
\email{n.sooppy@psau.edu.sa}
\affiliation{Department of Mathematics, College of Science and Humanities in Al-Kharj, Prince Sattam Bin Abdulaziz University, Al-Kharj 11942, Saudi Arabia}

\author{S. G. Nana Engo}
\email{serge.nana-engo@facsciences-uy1.cm}
\affiliation{Department of Physics, Faculty of Science, University of Yaounde I, P.O.Box 812, Yaounde, Cameroon}

\begin{abstract}
We propose a theoretical scheme to generate nonreciprocal bipartite entanglement between a cavity mode and vibrational modes in a molecular cavity optomechanical system. Our system consists of $\mathcal{N}$ molecules placed inside a spinning whispering-gallery-mode (WGM) resonator. The vibrational modes of these molecules are coupled to the WGM resonator mode (which is analogous to a plasmonic cavity) and the resonator is also coupled to an auxiliary optical cavity. We demonstrate that nonreciprocal photon-vibration entanglement and nonreciprocal vibration-vibration entanglement can be generated in this system, even at high temperatures. These nonreciprocal entanglements arise due to the Sagnac-Fizeau effect induced by the spinning WGM resonator. We find that spinning the WGM resonator in the counter-clockwise (CCW) direction enhances both types of nonreciprocal entanglement, especially under blue-detuned driving of the optical cavity mode. Furthermore, we show that vibration-vibration entanglement can be significantly enhanced by increasing the number of molecules. Our findings have potential applications in quantum information transmission and in the development of nonreciprocal quantum devices.
\end{abstract}

\maketitle

\section{\label{sec:Intro}Introduction} 

Investigating quantum effects at macroscopic scales is a central goal in modern physics, driving the exploration of various platforms that exhibit quantum behavior beyond the microscopic realm. These platforms range from superconducting circuits and trapped ions to optomechanical systems and hybrid quantum systems. Among these, cavity optomechanical systems, which explore the interaction between light and mechanical motion, have emerged as a versatile platform for studying fundamental quantum phenomena and developing novel quantum technologies \cite{Stannigel2012,Blais2020}. Within the realm of cavity optomechanics, significant attention has been directed toward enhancing light-matter interactions and achieving strong coupling regimes. Molecular cavity optomechanics, a burgeoning subfield, has emerged as a particularly promising avenue due to its unique characteristics and potential applications. This field, which encompasses the interaction between molecular polaritons (MPs) and optomechanics, has recently made notable progress within the scientific community\cite{Liu:21,Zhang2020}. Molecular cavity optomechanics broadens the scope of traditional optomechanics by accessing a regime characterized by ultra-high mechanical frequencies associated with molecular vibrations, offering a pathway to explore quantum phenomena at unprecedented scales. These systems provide a promising platform for investigating various physical phenomena, including Surface-Enhanced Raman Spectroscopy \cite{Roelli2015,Zou2024,Esteban2022}, Surface-Enhanced Frequency Conversion \cite{Roelli2020,Chen2021,Benda2022}, and polariton-like hybrid light-matter states \cite{Koner2023,Roelli2024}. In molecular optomechanical systems, molecular vibrations, often treated as mechanical oscillators, interact with confined optical fields. The interaction between a molecular vibration and a plasmonic mode, for example, is analogous to the interaction between cavity photons and a mechanical mode in traditional optomechanics \cite{Esteban2022}. These molecular vibrations are paramount, providing valuable information about molecular interactions with their environment and offering insights into their physical and chemical properties \cite{Pezacki2011}. Furthermore, coupling of optical fields to molecular vibrations can induce Raman scattering, a process that describes energy exchange between light and molecular vibrations through inelastic scattering within a quantum framework \cite{Roelli2024}. Cavity optomechanics has been effectively applied to study the behavior of interacting molecules in surface plasmon systems driven by two-color laser beams \cite{Liu:17}. The inherent interest in molecular cavity optomechanics stems from its ability to achieve higher vibrational frequencies and larger optomechanical coupling rates compared to macroscopic mechanical oscillators.

In light of the rapid progress in cavity optomechanical systems, nonreciprocal physics has witnessed a remarkable advancement in recent years. Due to the ability of nonreciprocal devices to allow light to propagate only in one direction (unidirectional) while blocking it in the opposite direction, they serve as interesting tools in building information processing networks~\cite{Flamini_2019,Shoji_2014}. Quantum entanglement has been achieved in a variety of optomechanical systems \cite{Vitali2007,Djorwe2024,Rostand2024,Rostand2025}, and they have stimulated a number of quantum technologies and sensing applications \cite{Djorwe2019,Tchounda2023,Djorwe2024}. Quantum entanglement has been studied in spinning cavity optomechanical systems, and they  have led to a strong correlation between photons and phonons in a particular direction (say clockwise), but weak correlation in the opposite direction (say counterclockwise). This results to nonreciprocal entanglement, which is mainly realized in cavity magnomechanics \cite{Chen2024,Chakraborty2023,liu2024twice} and optomagnomechanical systems~\cite{Zheng2024}. This nonreciprocal entanglement is often achieved by exploiting the Sagnac-Fizeau effect to induce an opposite frequency shift in the cavity~\cite{Chen2024,POST1967}. Nonreciprocal entanglement in cavity optomechanical systems has recently gained significant interest within the scientific world~\cite{Jiao2020,Ren:22,Zheng2023,Chen2023,Huang2024,Emale2025}. Moreover, coherent feedback control schemes have also been explored to enhance entanglement in various quantum systems, including cavity optomechanical systems and cavity magnomechanical systems, offering alternative routes to manipulate and amplify quantum correlations. For instance, a coherent feedback loop scheme has been proposed to enhance magnon-photon-phonon entanglement in cavity nanomechanics, demonstrating the potential of feedback mechanisms in quantum entanglement engineering \cite{Amazioug2023a}. These advancements highlight the growing interest in exploring and improving quantum entanglement in diverse physical systems for quantum information processing and other quantum technologies.

Quantum entanglement has been recently investigated in molecular optomechanical systems, and robust entanglement was generated owing to the strong optomechanical coupling resulting from the ultrasmall volume plasmonic cavities and the ultrahigh vibrational frequencies. A strong interaction between light-matter in hybrid photonic-plasmonic resonators\cite{Xiao2012}. This hybrid system consisted of a metal nanoparticle and a microcavity in the whispering gallery (WGM). Recently, whispering gallery mode (WGM) has garnered significant attention and the spinning resonator has been demonstrated in a recent experiment \cite{Maayani2018}. Quantum entanglement between vibrational modes can play a pivotal role in exploring spectroscopy and metrology. Such a vibration-vibration entanglement was studied recently in \cite{Huang2024a}.

Inspired by the aforementioned studies, we propose a theoretical scheme to generate and enhance nonreciprocal bipartite entanglement in a hybrid molecular cavity optomechanical system. In this system, the plasmonic cavity and $\mathcal{N}$ identical molecules are placed in a spinning WGM resonator driven by a pumping field and evanescently coupled to a tapered fiber. The spinning WGM resonator is coupled to an auxiliary cavity in a nonreciprocal manner. It is found that incorporating a plasmonic cavity and $\mathcal{N}$ molecules into a WGM resonator leverages the properties of the plasmonic mode, molecules and WGM resonator to enhance the light-matter interaction. Our findings reveal that bipartite cavity-vibration entanglement and vibration-vibration entanglement are significantly enhanced  under nonreciprocal conditions. The Sagnac effect induces noreciprocity in our system. Our proposed scheme put forward the following features: (i) realization of bipartite nonreciprocal entanglement, (ii) remarkable enhancement of vibration-vibration nonreciprocal entanglement, and (iii) robustness of bipartite entanglement against the decoherence effect. This proposed scheme may find applications in quantum information processing and the development of novel nonreciprocal devices.

The rest of the work is as follows. We give details of our model in \Cref{sec:Model}, together with the formulation of the Hamiltonian and dynamical equations. The nonreciprocal bipartite entanglement and the bidirectional contrast ratio are carried out in \Cref{sec:Resul}. Our conclusion is summarized in \Cref{sec:Concl}.

\section{\label{sec:Model}Model and dynamics}

\begin{figure}[htp!]
	\setlength{\lineskip}{0pt}
	\centering
	\includegraphics[width=.95\linewidth]{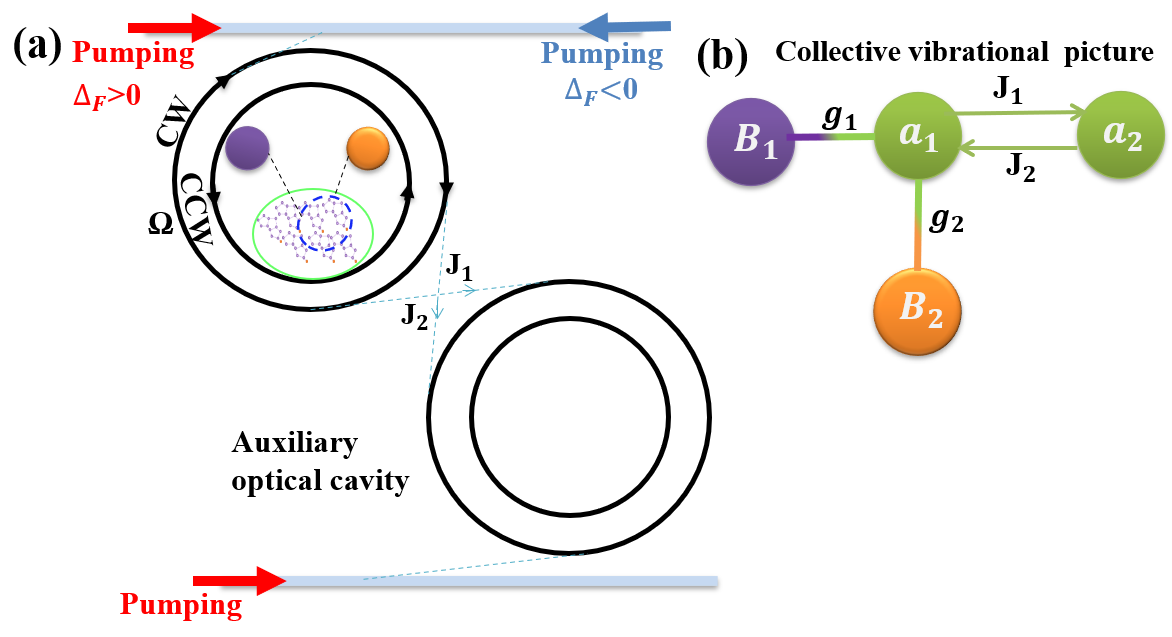}
    \caption{(a) Schematic of a molecular optomechanical system which consists of a spinning whispering gallery mode (WGM) and $\mathcal{N}$ molecules. This WGM is driven by the pumping field from left to right. The plasmonic cavity (represented by a mode $a_1$ within the WGM resonator) is coupled with the molecular vibration via optomechanical interaction and to an auxiliary optical cavity in a nonreciprocal manner. (b) Diagrammatic representation of the interaction among collective vibrational modes and cavity mode, which in turn interact with the auxiliary optical cavity. Note that $B_1$ and $B_2$ are collective vibrational modes derived from the same set of $\mathcal{N}$ molecular vibrations, not physically distinct sets of molecules. Their separation is a theoretical construct for simplifying analysis, as explained in the text.}
	\label{fig:Setup}
\end{figure}

In this section, we present a detailed description of our hybrid molecular cavity optomechanical system and derive the equations that govern its dynamics. As depicted in \Cref{fig:Setup}(a), the system comprises a plasmonic cavity and $\mathcal{N}$ molecules located within a spinning whispering-gallery mode (WGM) resonator. It should be noted that, in our model, the plasmonic cavity is not a separate physical cavity, but rather refers to the cavity mode within the WGM resonator (denoted as mode $a_1$) that interacts with the molecular vibrations and exhibits characteristics analogous to plasmonic modes of a nanoparticle, as justified by \textcite{Esteban2022}. The WGM resonator is evanescently coupled to a tapered fiber, facilitating the input and output of light. Furthermore, this spinning WGM resonator is coupled to an auxiliary optical cavity in a nonreciprocal manner. This nonreciprocal coupling is engineered through the rotation of the WGM resonator and is direction-dependent: when the WGM resonator rotates in the clockwise (CW) direction, the auxiliary optical cavity (mode $a_2$) couples to it with a strength $J_1$ in one direction; conversely, in the counter-clockwise (CCW) direction, the coupling occurs in the opposite direction with a strength $J_2$. This directionality, stemming from the Sagnac-Fizeau effect (discussed below), is key to achieving nonreciprocal entanglement. We emphasize that while \Cref{fig:Setup}(a) schematically shows separate cavities for illustration, the plasmonic cavity in our model is embodied by the WGM resonator mode $a_1$. \Cref{fig:Setup}(b) provides a diagrammatic representation of the interactions between the collective vibrational modes and the cavity modes. Both the spinning WGM resonator (mode $a_1$) and the auxiliary optical cavity (mode $a_2$) are driven by strong pumping fields with amplitudes $\mathcal{E}_1$ and $\mathcal{E}_2$, respectively. For simplicity and without loss of generality, we assume these amplitudes to be equal, i.e., $\mathcal{E}_1 = \mathcal{E}_2 = \mathcal{E}$. The molecular vibrations are modeled as harmonic oscillators with a uniform frequency $\omega_m$.

\subsection{Hamiltonian of the system}\label{sec:AII}

The Hamiltonian of the system under consideration takes the form,
\begin{equation}\label{eq:1}
\begin{aligned}
H=&\Delta a_1^\dagger a_1+\Delta_{c2} a_2^\dagger a_2  +\sum_{j=1}^\mathcal{N} \omega_m b_j^\dagger b_j+\sum_{j=1}^\mathcal{N} g_m a_1^\dagger a_1(b_j^\dagger + b_j)+J_1a_1^\dagger a_2+J_2a_2^\dagger a_1 +  i\mathcal{E}_1(a_1^\dagger - a_1)+i\mathcal{E}_2(a_2^\dagger - a_2)
\end{aligned}
\end{equation}
where $a_1(a_1^\dagger)$, $a_2(a_2^\dagger)$ and $b_j(b_j^\dagger)$ are the annihilation (creation) operators of the WGM resonator mode (analogous to the plasmonic cavity mode), auxiliary optical cavity mode and the $j^{th}$ molecular mode, respectively. In our model, $a_1$ represents a single effective mode of the WGM resonator relevant to the interactions. The nonreciprocal coupling between $a_1$ and the auxiliary cavity mode $a_2$ (represented by the terms $J_1 a_1^\dagger a_2$ and $J_2 a_2^\dagger a_1$) arises from the Sagnac-Fizeau effect on the spinning WGM resonator, which makes the effective coupling strength dependent on the direction of energy transfer between the modes. In the above Hamiltonian, the quantity $\Delta_{cj}=\omega_{cj}-\omega_{L}$ ($j=1,2$) represents the frequency detunings that is introduced by moving in the frame rotating at the driving frequency $\omega_{L}$, where $\omega_{cj}$ are the resonance frequencies of the effective WGM mode ($a_1$) and auxiliary cavity modes. We have introduced  $\Delta=\Delta_{c1}-\Delta_F$, where $\Delta_F$ stands for the Sagnac-Fizeau shift of the resonance frequency of the cavity induced by the circulating light in the spinning WGM resonator. This Sagnac-Fizeau shift is the physical origin of the nonreciprocal coupling in our system, distinguishing $J_1$ and $J_2$. The first three terms of our Hamiltonian capture the free energies of the two cavities and for the molecular vibrational modes. The fourth term represents the optomechanical coupling between the WGM resonator mode ($a_1$) and the molecular vibrational mode through the coupling strength $g_m$. The fifth and sixth terms refer to the nonreciprocal coupling between the WGM resonator mode ($a_1$) and auxiliary cavity with strengths $J_1$ and $J_2$. The last two terms denote the driving of the WGM resonator mode ($a_1$) and the auxiliary cavity by the pumping field. The resonance frequency of the   vibrational modes is captured by $\omega_m$. If the WGM resonator is assumed to be rotated in the CW sense with the angular velocity $\Omega$, the Sagnac-Fizeau shifts take the form\cite{RevModPhys.39.475},
\begin{equation}\label{eq:2}
\Delta_F=\pm\frac{n\Omega R\omega_{c1}}{c}\left(1-\frac{1}{n^2}-\frac{\lambda}{n}\frac{dn}{d\lambda}\right),
\end{equation}
where, $n$ is the refractive index, R is the radius of the resonator, $\lambda$ is the wavelength of light in a vacuum and c is speed of light in a vacuum. The dispersion term $\frac{dn}{d\lambda}$, can be ignored since it is relatively small. The plus and minus sign in \eqref{eq:2} corresponds to CW and CCW driving. Since we assumed that the cavity is rotating in the CW sense, then we consider $\Delta_F>0$ ($\Delta_F<0$) for CW (CCW)  pumping field, respectively. in our investigation, we will assume a large number of molecules in the cavity, and  we define the two collective vibrational modes,
\begin{equation}\label{eq:3}
B_1=\sum_{j=1}^{M}b_j/\sqrt{M}, ~~~B_2=\sum_{j=M+1}^{\mathcal{N}}b_j/\sqrt{\mathcal{N}-M}.
\end{equation}
This separation into two collective modes, $B_1$ and $B_2$, is a theoretical construct for simplifying the analysis and does not imply that the molecules are physically separated into two distinct groups. Instead, it allows us to distinguish the collective vibrations that are more strongly coupled to the WGM resonator mode ($a_1$) from the rest, for the purpose of analyzing entanglement properties.
By using these modes, our Hamiltonian can be rewritten as,
\begin{equation}\label{eq:4}
\begin{aligned}
H=&\Delta a_1^\dagger a_1+\Delta_{c2} a_2^\dagger a_2  +\sum_{k=1}^2\left(\omega_m B_k^\dagger B_k+g_k a_1^\dagger a_1(B_k^\dagger + B_k)\right)+J_1a_1^\dagger a_2+J_2a_2^\dagger a_1 +  i\mathcal{E}_1(a_1^\dagger - a_1)+i\mathcal{E}_2(a_2^\dagger - a_2),
\end{aligned}
\end{equation}
 where $g_1=g_m\sqrt{M}$ and $g_2=g_m\sqrt{\mathcal{N}-M}$ represent the collective optomechanical coupling strength between the WGM resonator mode ($a_1$) and the collective vibrational mode $B_1(B_2)$. The quantities M and $\mathcal{N}$ are the distribution number of molecular collective mode and number of molecules, respectively.
 
 \subsection{\label{sec:IIB}Quantum Langevin equations (QLE)} 
 
The dynamical state of our system is described by the following Quantum Langevin Equations (QLEs),
\begin{equation}\label{eq:5}
\begin{aligned}
\dot{a}_1=&-(i\Delta+\kappa_1)a_1-ia_1\sum_{k=1}^{2}g_k(B_k+B_k^\dagger)-iJ_1a_2+\mathcal{E}_1+\sqrt{2\kappa_1}a_1^{\text in},\\
\dot{a}_2=&-(i\Delta_{2c}+\kappa_2)a_2-iJ_2a_1+\mathcal{E}_2+\sqrt{2\kappa_2}a_2^{\text in},\\
\dot{B}_1=&-(i\omega_m+\gamma_1)B_1-ig_1a_1^\dagger a_1+\sqrt{2\gamma_1}B_1^{\text in},\\
\dot{B}_2=&-(i\omega_m+\gamma_2)B_2-ig_2a_1^\dagger a_1+\sqrt{2\gamma_2}B_2^{\text in},
\end{aligned}
\end{equation} 
 where the related noise operators $\mathcal{O}^{\text{in}} (\mathcal{O}=a_j,B_j)$ have been taken into account. We would like to mention that,
 \begin{align}
 &B^{\text{in}}_1=\sum_{j=1}^{M}\frac{b^{\text{in}}_j}{\sqrt{M}},
 &B^{\text{in}}_2=\sum_{j=M+1}^{\mathcal{N}}B^{\text{in}}_1=\sum_{j=1}^{M}\frac{b^{\text{in}}_j}{\sqrt{\mathcal{N}-M}}.
 \end{align} 
 These noise operators have zero mean values, and are characterized by the correlation functions,
 \begin{equation}\label{eq:6}
 \begin{aligned}
 &\langle a_k^{\text{in}}(t)a_k^{\text{in}\dagger}(t^\prime)\rangle=\delta(t-t^\prime),\\
 &\langle B_k^{\text{in}}(t)B_k^{\text{in}\dagger}(t^\prime)\rangle=(n_k+1)\delta(t-t^\prime),\\
 &\langle B_k^{\text{in}\dagger}(t)B_k^{\text{in}}(t^\prime)\rangle=n_k\delta(t-t^\prime),
 \end{aligned}
 \end{equation}
 where $n_j=\left\{\exp\left(\frac{\hbar\omega_j}{k_B\text{T}}\right)-1\right\}^{-1}$ denotes the thermal phonon number at temperature $T$ and $k_B$ is the Boltzmann constant. We assume that $\gamma_1$=$\gamma_2$=$\gamma$ for the two collective vibrational modes and $\kappa_1$=$\kappa_2$=$\kappa$ for the cavity modes. For the cavity modes, the thermal excitation number can be neglected in the optical frequency band.
 
 The equations given in \eqref{eq:5} involve nonlinearities, and they need to be linearized  in order to capture their fluctuation dynamics. For this purpose, each operator ( $\mathcal{O}$) is split into their mean value ($\langle\mathcal{O}\rangle$) plus small fluctuations ($\delta\mathcal{O}$) around them, i.e., $\mathcal{O}=\langle\mathcal{O}\rangle+\delta\mathcal{O}$, where $\mathcal{O}\equiv a_k,B_k$, and $\langle\mathcal{O}\rangle\equiv\alpha_k, \beta_k$. For long term behavior, the mean value equations are no longer time dependent, and they reach the following steady state equations,
 \begin{equation}\label{eq:7}
 \begin{aligned}
 \alpha_1=&\frac{\mathcal{E}_1-iJ_1\alpha_2}{(i\Delta^\prime+\kappa_1)},\\
 \alpha_2=&\frac{\mathcal{E}_2-iJ_2\alpha_1}{(i\Delta_2+\kappa_2)},\\
 \beta_k=&\frac{-ig_k|\alpha_k|^2}{(i\omega_m+\gamma_k)}.\\
 \end{aligned}
 \end{equation}
 Similarly, we derived the set of equations for the fluctuation operators,
 \begin{equation}\label{eq:9}
 \begin{aligned}
 \delta\dot{a}_1=&-(i\Delta^\prime+\kappa_1)\delta a_1-i\sum_{k=1}^{2}G_k (\delta B_k+\delta B_k^\dagger)\\&-iJ_1\delta a_2+\sqrt{2\kappa_1}a_1^{\text in},\\
 \delta\dot{a}_2=&-(i\Delta_2+\kappa_2)\delta a_2-iJ_2\delta a_1+\sqrt{2\kappa_2}a_2^{\text in},\\
 \delta\dot{B}_1=&-(i\omega_m+\gamma_1)\delta B_1-i(G_1^\ast\delta a_1+G_1\delta a_1^\dagger)+\sqrt{2\gamma_1}B_1^{\text in},\\
 \delta\dot{B}_2=&-(i\omega_m+\gamma_2)\delta B_2-i(G_2^\ast\delta a_1+G_2\delta a_1^\dagger)+\sqrt{2\gamma_2}B_2^{\text in},\\
 \end{aligned}
 \end{equation}
 where $\Delta^\prime=\Delta+\sum_{k=1}^{2}2g_k\text{Re}[\beta_k]$ is the effective detuning, $G_1=\sqrt{M}g_m\alpha_1$, and $G_2=\sqrt{\mathcal{N}-M}g_m\alpha_1$, are the effective optomechanical coupling strengths.
 
In order to study the entanglement, we define the following quadrature operators, 
\begin{equation}
\begin{aligned}
\delta x_k&=\frac{( \delta a_k+\delta a_k^{\dagger})}{\sqrt{2}}, \delta y_k=\frac{( \delta a_k- \delta a_k^{\dagger})}{i\sqrt{2}},\\
\delta q_k&=\frac{(\delta B_k+\delta B_k^{\dagger})}{\sqrt{2}},\delta p_k=\frac{(\delta B_k-\delta B_k^{\dagger})}{i\sqrt{2}},
\end{aligned}
\end{equation}
with $k=1,2$. The corresponding quadrature  noise operators are,
\begin{equation}
\begin{aligned}
x_k^{\text{in}}&=\frac{(a_j^{\text{in}}+ a_k^{\text{\text{in}}\dagger})}{\sqrt{2}},~ y_k^{\text{in}}=\frac{( a_k^{\text{\text{in}}}- a_k^{in\dagger})}{i\sqrt{2}},\\
q^{\text{in}}_k&=\frac{( B_k^{\text{in}}+ B_k^{\text{in}\dagger})}{\sqrt{2}},~ p^{\text in}_k=\frac{( B_k^{\text{in}}- B_k^{in\dagger})}{i\sqrt{2}}.
\end{aligned}
\end{equation}
By using these quadratures operators, our linearized equations displayed in \eqref{eq:9} can be put into the following form,
\begin{equation}\label{eq:14}
\dot{\chi}(t)=A\chi(t)+n(t),
\end{equation} 
 where A is $8\times 8$ matrix which reads,
 \begin{widetext}
 \begin{equation}
 A=
 \begin{pmatrix}
 -\kappa_1&\Delta^\prime&0&J_1&2\text{Im}[G_1]&0&2\text{Im}[G_2]&0\\
 -\Delta^\prime&-\kappa_1&-J_1&0&-2\text{Re}[G_1]&0&-2\text{Re}[G_2]&0\\
 0&J_2&-\kappa_2&\Delta_{2c}&0&0&0&0\\
 -J_2&0&-\Delta_{2c}&-\kappa_2&0&0&0&0\\
 0&0&0&0&-\gamma_1&\omega_m&0&0\\
 -2\text{Re}[G_1]&-2\text{Im}[G_1]&0&0&-\omega_m&-\gamma_1&0&0\\
 0&0&0&0&0&0&-\gamma_2&\omega_m\\
 -2\text{Re}[G_2]&-2\text{Im}[G_2]&0&0&0&0&-\omega_m&-\gamma_2\\
 \end{pmatrix}.
 \end{equation}
 \end{widetext}
 The vectors of the quadrature fluctuation operators are defined by
 \begin{widetext}
 	\begin{equation}
 	\begin{aligned}
 	&\chi^{T}(t)=\left(\delta x_1(t),\delta y_1(t),\delta x_2(t),\delta y_2(t),\delta q_1(t),\delta p_1(t),\delta q_2(t),\delta p_2(t)\right),\\ &N^{T}(t)=\left(\sqrt{2\kappa}x_1^{\text{in}},\sqrt{2\kappa}y_1^{\text{in}},\sqrt{2\kappa}x_2^{\text{in}},\sqrt{2\kappa}y_2^{\text{in}},\sqrt{2\gamma_1}q_1^{\text{in}},\sqrt{2\gamma_1}p_1^{\text{in}},\sqrt{2\gamma_2}q_2^{\text{in}}, \sqrt{2\gamma_2}p_2^{\text{in}}\right).
 	\end{aligned}
 	\end{equation}
 \end{widetext}
 
 Our dynamical equations are meaningless if our system is not stable. Therefore, we use linear stability theory to analyzes the stability of our system, i.e., all real parts of the eigenvalues of the drift matrix A must be negative. This stability requirement follows the Routh-Hurwitz criterion \cite{DeJesus1987}. Under these conditions,  our system can be fully characterized by the $8\times8$ covariance matrix (CM), which is obtained by solving the following Lyapunov equation,
 \begin{equation}
 A{V}+VA^{T}=-D,
 \end{equation}
 where $V_{ij}=\frac{1}{2}\left\{\chi_{i}(t)\chi_j(t^{\prime})+\chi_j(t^{\prime})\chi_{i}(t)\right\}$, the diffusion matrix is given by,
 \begin{widetext} 
 \begin{equation}
 D=\text{diag}\left[\kappa_1,\kappa_1,\kappa_2,\kappa_2,\gamma_1(2n_{\text{B1}}+1),\gamma_1(2n_{\text{B1}}+1),\gamma_2(2n_{\text{B2}}+1),\gamma_2(2n_{\text{B2}}+1)\right],
 \end{equation}
\end{widetext}
 and is defined through $\langle v_{i}(t)v_j(t^\prime)+v_j(t^\prime)v_{i}(t)\rangle/2=D_{ij}\delta(t-t^\prime)$.
 
From the covariance matrix, we quantify the quantum entanglement in our system by using the logarithmic negativity $E_N$. The logarithmic negativity is expressed as \cite{PhysRevLett.95.090503},
\begin{equation}
E_N=\text{max}\left[0,-\text{ln}(2\zeta)\right],
\end{equation}
where $\zeta\equiv2^{-1/2}\left\{\sum(V)-\left[\sum(V)^{2}-4\text{det}({V})\right]^{1/2}\right\}^{1/2}$, with $\sum(V)=\text{det}(\mu_1)+\text{det}(\mu_2)-2\text{det}(\mu_3)$. 
The $\mu_k$ are the elements of the considered bipartite submatrix extracted from the covariance matrix of the whole system. For a given subsystem, consisting for instance of modes $1$ and $2$, the corresponding submatrix is,
\begin{equation}
V_{sub}=
\begin{pmatrix}
\mu_1&\mu_3\\
\mu_3^{T}&\mu_2\\
\end{pmatrix}.
\end{equation} 
 
\section{\label{sec:Resul}Results and discussion}

In this section, we present and discuss the results of our numerical analysis, focusing on the stability of the proposed system and the characteristics of nonreciprocal entanglement. To numerically simulate the system behavior and quantify the entanglement, we employ experimentally feasible parameters drawn from the relevant literature on molecular optomechanics and related systems \cite{Xiao2012,Liu:17,Zou2024,Schmidt_2024}: molecular vibrational frequency $\omega_m/2\pi=\qty{30}{\tera\hertz}$, optomechanical coupling strength $g_m/2\pi=\qty{30}{\giga\hertz}$, cavity decay rate $\kappa_1/\omega_m=0.3$, $\kappa_2=\kappa_1$, cavity detuning $\Delta_{1c}=\omega_m$, $\Delta_{2c}=\omega_m$, vibrational decay rate $\gamma_1/\omega_m=10^{-4}$, $\gamma_2=\gamma_1$, driving amplitude $\mathcal{E}/\omega_m=16$, coupling strengths $J_1/\omega_m=0.3$, $J_2/\omega_m=1$, molecular distribution number $M=\num{50}$, number of molecules $\mathcal{N}=\num{100}$, temperature T=\qty{312}{\kelvin}, Sagnac-Fizeau shift magnitude $|\Delta_F|=0.1\omega_m$ \cite{Chen2024}. These parameters are chosen to ensure that our theoretical predictions are grounded in experimentally achievable regimes and allow for observable quantum effects.

\subsection{Stability analysis} 

\begin{figure}[htp!]
	\centering
	\includegraphics[width=1.05\linewidth]{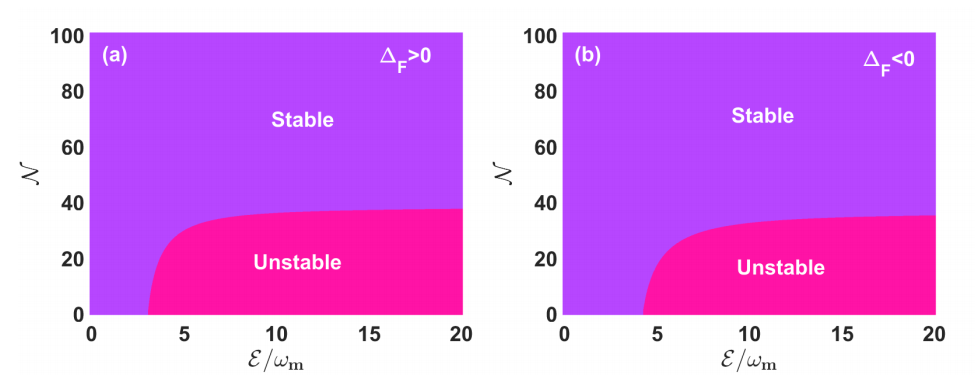}
	\caption{The dependence of the system stability on the driving amplitude $\mathcal{E}$ and number of molecules $\mathcal{N}$. (a) corresponds to $\Delta_F>0$, and (b) corresponds to $\Delta_F<0$. Parameters are chosen as: $\omega_m/2\pi=\qty{30}{\tera\hertz}$, $g_m/2\pi=\qty{30}{\giga\hertz}$, $\kappa_1/\omega_m=0.3$, $\kappa_2=\kappa_1$, $\Delta_{1c}=\omega_m$, $\Delta_{2c}=\omega_m$, $\gamma_1/\omega_m=10^{-4}$, $\gamma_2=\gamma_1$, $\mathcal{E}/\omega_m$=16, $J_1/\omega_m=0.3$, $J_2/\omega_m=1$, M=50, $\mathcal{N}=100$, T=\qty{312}{\kelvin} and $|\Delta_F|=0.1\omega_m$}
	\label{fig:2}
\end{figure}

We have displayed in \Cref{fig:2} the stability analysis of our system as a function of the driving amplitude $\mathcal{E}$ and the number of molecules $\mathcal{N}$. These figures correspond to $\Delta_F>0$ (\Cref{fig:2}a) and to $\Delta_F<0$ (\Cref{fig:2}b).
 
For each figure, the stable region is shown in purple color, while the unstable area is marked in pink. This stability is a prerequisite for realizing steady-state quantum entanglement in our system. The stability of the system is determined using the traditional Routh-Hurwitz criterion. As illustrated in \Cref{fig:2}, our numerical results reveal that our system is unstable only within the region $3.16\leq\mathcal{E}/\omega_m\leq20$ and $0\leq\mathcal{N}\leq30$ for CW spinning and $4.26\leq\mathcal{E}/\omega_m\leq20$ and $0\leq\mathcal{N}\leq30$ for CCW spinning of the WGM resonator. This indicates that for lower driving amplitudes or a higher number of molecules, the system operates in a stable regime, which is essential to observe steady-state entanglement. In contrast, a stronger driving and fewer molecules can lead to instability, which must be avoided in experimental implementations. This suggests that a large number of molecules in the cavity ($\mathcal{N}\geq40$) induce stability regardless of driving strength, while a weak number of molecules ($\mathcal{N}\leq30$) trigger instability as the driving amplitude becomes strong enough. The result is that to maintain the stability of the system under the condition of a strong driving amplitude, it is necessary to increase the number of molecules. The parameters we will use from now on fulfill this stability condition.

 \subsection{Nonreciprocal entanglement analysis} 
 
To explore nonreciprocal entanglement in our system, the logarithmic negativity is used to quantify bipartite entanglement accompanied by the Sagnac-Fizeau effect. The quantum entanglement between the modes $a_2, B_1$ and $B_1, B_2$ is represented by the logarithmic negativity $E_{a2B1}$ and $E_{B1B2}$ respectively. Due to the fact that auxiliary cavity-vibration entanglements $E_{a2B1}$ and $E_{a2B2}$  exhibit similar features, we will discuss only $E_{a2B1}$ for simplicity.

 \begin{figure*}[htp!]
 	\centering
 	\includegraphics[width=0.9\linewidth]{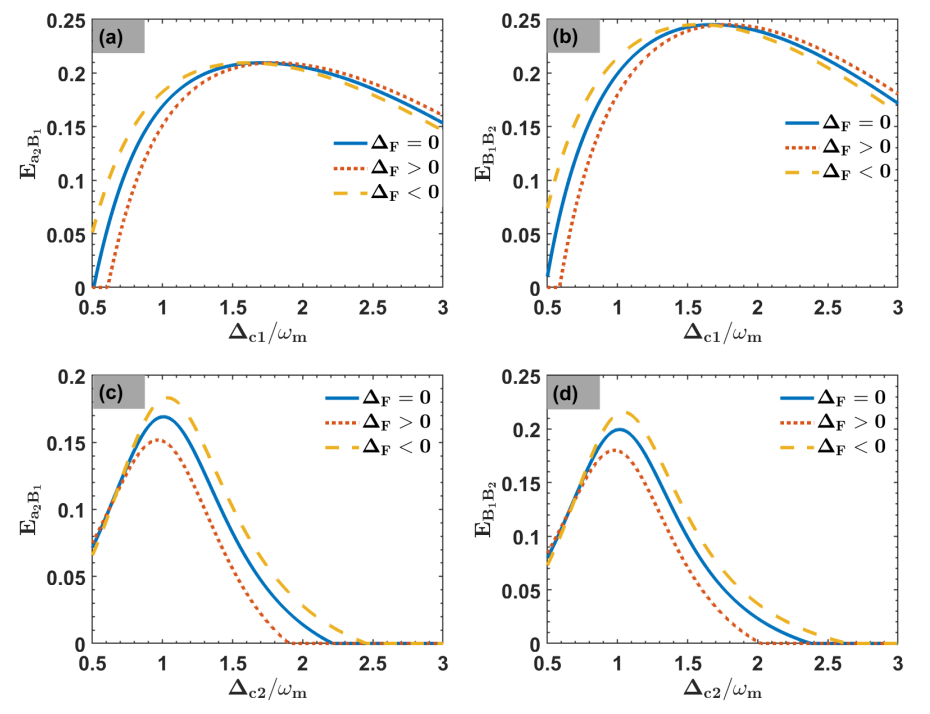}
    \caption{(a) Photon-vibration entanglement versus the normalized cavity detuning $\Delta_{1c}/\omega_m$, (b) vibration-vibration entanglement versus the normalized cavity detuning $\Delta_{1c}/\omega_m$, (c) Photon-vibration entanglement versus the normalized auxiliary cavity detuning $\Delta_{2c}/\omega_m$, and (d) vibration-vibration entanglement versus the normalized auxiliary cavity detuning $\Delta_{2c}/\omega_m$. Parameters are chosen as: $\Delta_F/\omega_m=0$ (Azure blue--solid line), $\Delta_F/\omega_m>0$ (dark red--dotted line), $\Delta_F/\omega_m<0$ (golden yellow--dashed line). Other parameters are the same as those in \Cref{fig:2} and listed at the beginning of \Cref{sec:Resul}.}
 	\label{fig:3}
 \end{figure*}

It can be seen from \Cref{fig:3} that photon-vibration entanglement and vibration-vibration entanglement are tuned by varying frequency detunings $\Delta_{1c}$ and $\Delta_{2c}$. The optimal entanglement is obtained when the WGM resonator is rotating in the CCW direction at $\Delta_{1c}\approx1.5\omega_m$ and $\Delta_{2c}\approx\omega_m$. This shows that optimal entanglement is generated in our system when it is tuned in the blue sideband parameter regime ($\Delta_F<0$) where the counter rotating wave term plays a dominant role. This observation aligns with previous studies on nonreciprocal entanglement in spinning optomechanical systems, where blue-detuned driving and CCW rotation are found to enhance nonreciprocity \cite{Chen2024}. On the other hand, by driving the system with a red-detuned laser field ($\Delta_F>0$), the molecular vibrations can be cooled to the ground state. This cooling of the molecular vibration mode is a prerequisite for the preparation of an entangled state between molecules and other degrees of freedom. Moreover, we expect the nonreciprocal vibration-vibration mode in our system to be enhanced by the Sagnac-Fizeau effect.

 \begin{figure}[htp!]
 	\centering
 	\includegraphics[width=1.05\linewidth]{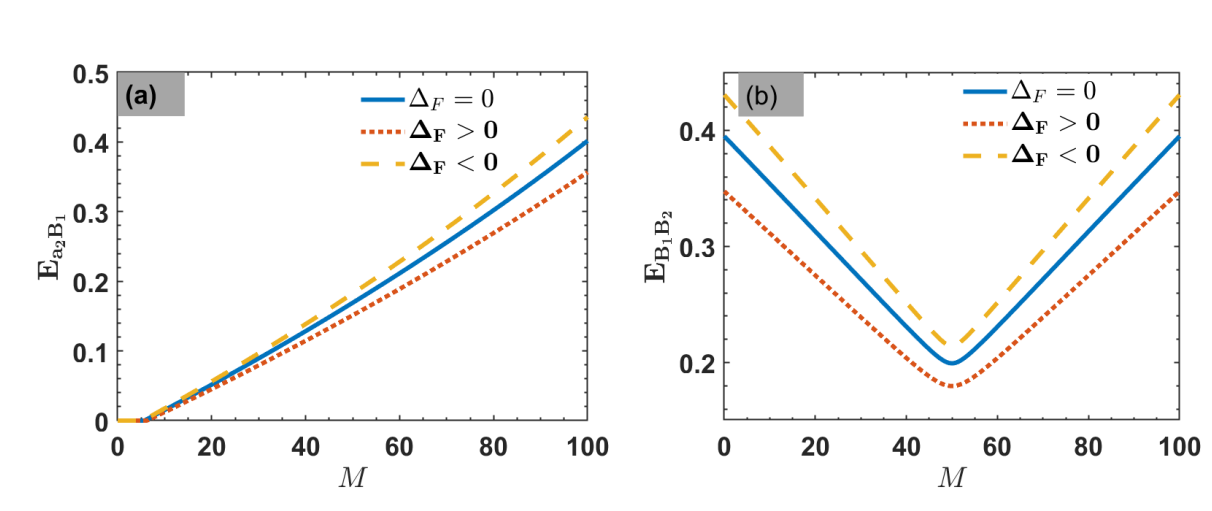}
    \caption{(a) Photon-vibration entanglement versus the distribution number of molecular collective mode M, and (b) vibration-vibration entanglement versus the distribution number of molecular collective mode $M$. Parameters are chosen as: $\Delta_F/\omega_m=0$ (Azure blue--solid line), $\Delta_F/\omega_m>0$ (dark red--dotted line), $\Delta_F/\omega_m<0$ (golden yellow--dashed line). Other parameters are the same as those in \Cref{fig:2} and listed at the beginning of \Cref{sec:Resul}.}
 	\label{fig:4}
 \end{figure}

\Cref{fig:4} depicts the influence of the number of molecular collective mode distributions on entanglement. It can be seen that photon-vibration entanglement increases as the number of molecular collective modes $M$ increases (see \Cref{fig:4}a). On the other hand, the vibration-vibration entanglement decreases at first and reaches its minimum for $M=\mathcal{N}$/2, where it starts to grow as $M$ increases (see \Cref{fig:4}b).

\begin{figure*}[htp!]
	\centering
	\includegraphics[width=1\linewidth]{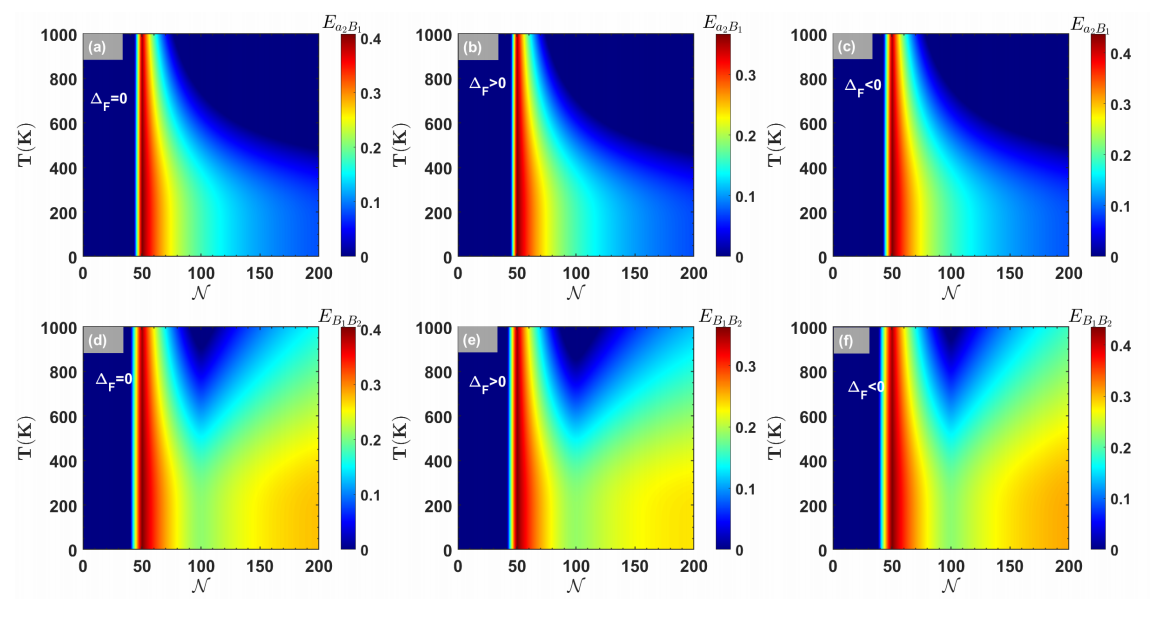}
	\caption{Panels (a)--(c); contour plot of photon-vibration entanglement $E_{a_2B_1}$  versus the molecular number $\mathcal{N}$ and the bath temperature $T$ for (a) $\Delta_F=0$, (b) $\Delta_F>0$, and (c) $\Delta_F<0$. Panels (d)--(f); contour plot of vibration-vibration entanglement $E_{B_1B_2}$versus the molecular number for (d) $\Delta_F$=0, (e) $\Delta_F>0$, and $\Delta_F<0$. Other parameters are the same as those in \Cref{fig:2} and listed at the beginning of \Cref{sec:Resul}. Note the color scale representing entanglement magnitude on the right of each column.}
	\label{fig:5}
\end{figure*}

\Cref{fig:5} displays the influence of temperature and number of molecules on entanglement. It can be seen that our system is robust enough against thermal decoherence, since our generated entanglement persists at temperatures relatively high. From these figures, one observes that the vibration-vibration entanglement persists more against thermal fluctuations as the molecular number $\mathcal{N}$ increases than the photon-vibration entanglement. In fact, in \Cref{fig:5}(a,b,c), the photon-vibration entanglement decreases significantly for $\mathcal{N}\gtrsim 150$. However, \Cref{fig:5}(d,e,f), display strong entanglement above $\mathcal{N}$=150. This contrasting behavior suggests that photon-vibration entanglement is more susceptible to thermal noise and increasing molecular number beyond a certain point, while vibration-vibration entanglement benefits from larger molecular ensembles and maintains robustness at higher temperatures. The observed robustness of vibration-vibration entanglement to temperature, particularly for larger $\mathcal{N}$, is a notable feature, potentially exceeding the thermal noise resilience seen in some other optomechanical entanglement schemes and highlighting the advantage of collective molecular vibrations for robust quantum correlations. Therefore, vibration-vibration entanglement is enhanced as the number of molecules increases in the system, while a large number of molecules destroys the photon-vibration entanglement. For a weak number of molecules ($\mathcal{N}\lesssim 50$), both types of entanglements. In fact, there is a sort threshold from which the entanglement is generated in our system, depending on $\mathcal{N}$. This phenomenon can be clearly seen in \Cref{fig:6}a which displays both entanglements versus the molecular number $\mathcal{N}$ at $T=\qty{200}{\kelvin}$, and $\Delta_F=0$ (the cases for $\Delta_F>0$ and $\Delta_F<0$ are similar). The plots on this figure are extracted from \Cref{fig:5}a (for the photon-vibration entanglement) and from \Cref{fig:5}d (for the vibration-vibration entanglement). As predicted before, it can be seen that both entanglements reach their threshold around $\mathcal{N}=50$ where they start to grow. This threshold is inferred from the instability limit shown in the stability diagram (see \Cref{fig:2}). Indeed, for a weak molecular number, the system is unstable and no entanglement is generated. As the system crosses the stable area, entanglement is generated and it exponentially increases at this threshold from where it starts to decrease, as depicted in \Cref{fig:6}a. Moreover, as the molecular number increases, it is observed that $E_{\alpha_2,B_1}$ decreases monotonically, while $E_{B_1,B_2}$ begins to increase. This indicates a trade-off, i.e., photon-vibration entanglement is stronger for fewer molecules but less robust, while vibration-vibration entanglement becomes dominant and more robust for larger molecular ensembles. This behavior could be attributed to the collective enhancement of vibration-vibration interactions with increasing molecule number, whereas the photon-vibration entanglement may be limited by saturation effects or increased complexity in the system at higher densities. This analysis shows how the vibration-vibration entanglement is robust as the number of molecules grows in our system. This suggests the generation of macroscopic entanglement in such a system, which may be interesting for quantum processing information and a number of quantum computation tasks. Furthermore, the observed robustness of vibration-vibration entanglement in a molecular system may offer advantages compared to systems relying on macroscopic mechanical oscillators, which can be more susceptible to environmental noise.

\subsection{Switchable nonreciprocity} 

 In this section, we use the following bidirectional contrast ration $\mathcal{C}$ for bipartite entanglement to quantitatively describe nonreciprocal entanglement \cite{Chen2023}, 
 \begin{equation}\label{eq:10}
 \mathcal{C}_{kl}=\frac{|E_{kl}(\Delta_F>0)-E_{kl}(\Delta_F<0)|}{E_{kl}(\Delta_F>0)+E_{kl}(\Delta_F<0)}.
 \end{equation}
The bidirectional contrast ration defined in \eqref{eq:10} satisfies the condition $0\leq\mathcal{C}\leq1$. The contrast ratio $\mathcal{C}$ is directly proportional to nonreciprocal  entanglement, i.e., the higher the contrast ratio $\mathcal{C}$, the higher is the nonreciprocal entanglement.

\begin{figure}[htp!]
	\centering
	\includegraphics[width=1.05\linewidth]{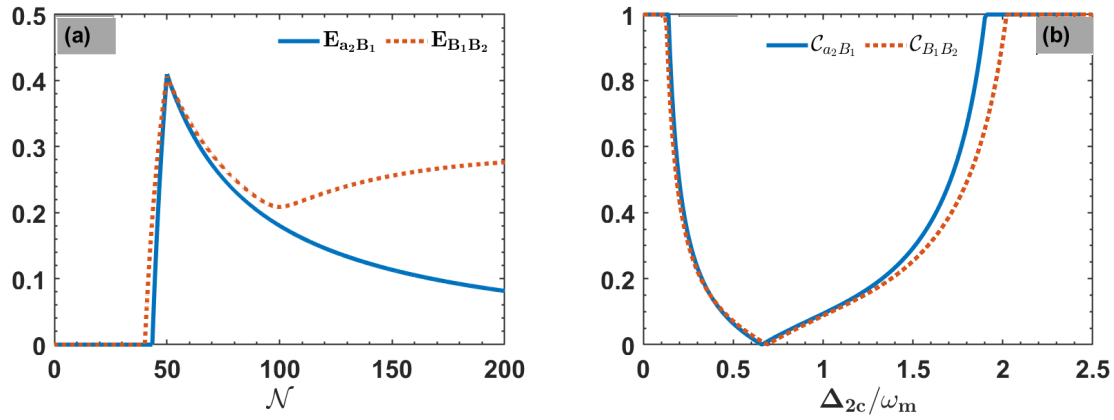}
	\caption{ (a) Bidirectional contrast ratio $\mathcal{C}$ for two types of bipartite entanglement as a function of auxiliary cavity detuning $\Delta_{2c}$; $\mathcal{C}_{a_2B_1}$ (Azure blue--solid line) and $\mathcal{C}_{B_1B_2}$ (dark red--dotted line) (b) Photon--vibration $E_{a_2B_1}$ (Azure blue--solid line) and vibration--vibration $E_{B_1B_2}$ (dark red--dotted line) bipartite entanglements versus the number of molecules $\mathcal{N}$ for $\Delta_F=0$ and T=\qty{200}{\kelvin}. Other parameters are the same as those in \Cref{fig:2} and listed at the beginning of \Cref{sec:Resul}}.
	\label{fig:6}
\end{figure}

We now use numerical analysis to study the influence of physical parameters on nonreciprocal entanglement. It is worth mentioning here that $\mathcal{C}_{kl}=0$ ($\mathcal{C}_{kl}=1$) corresponds to no (ideal) nonreciprocity for bipartite entanglement. One can see in \Cref{fig:6}(b) that nonreciprocity of bipartite entanglement can be switched on and off by adjusting the auxiliary cavity detuning $\Delta_{2c}$, i.e., the bidirectional contrast ratio for the two types of entanglement can be adjusted between $0$ and $1$ by varying $\Delta_{2c}$. Indeed, \Cref{fig:6}(b) shows how by adjusting the detuning of the auxiliary cavity frequency, it is possible to achieve optimal nonreciprocal photon--vibration and vibration--vibration entanglements. The optimal nonreciprocity is achieved in our scheme when $0<\Delta_{2c}/\omega_m<0.5$ and $1.5\leq\Delta_{2c}/\omega_m\leq2.5$. This indicates that nonreciprocal entanglement is achievable within a specific range of auxiliary cavity detunings, offering tunability for experimental realization. The tunability of nonreciprocity via cavity detuning, as demonstrated here, contrasts with some nonreciprocal devices where nonreciprocity is a fixed characteristic and offers an added degree of control for quantum applications. This shows that as the frequency detuning increases ($\Delta_{2c}>1.5\omega_m$), the entangled states are becoming more distinct and asymmetric in their behavior in different directions. However, nonreciprocity vanishes at the vicinity of $\Delta_{2c}/\omega_m\approx0.68$, i.e., the entangled states are indistinguishable in both directions. This possibility of tuning on and off these two types of entanglement through the bidirectional contrast ratio means that there is nonreciprocal process in the generation of the entanglement in our system. This tunability of the entanglement on demand is interesting for a number of quantum applications. Furthermore, the ability to switch nonreciprocity by tuning the auxiliary cavity detuning provides a valuable control knob for quantum information processing and nonreciprocal quantum device applications, potentially enabling on-demand routing or isolation of quantum signals.

\section{\label{sec:Concl}Conclusion}

In summary, we have proposed and theoretically analyzed a scheme to achieve nonreciprocal bipartite entanglement in a hybrid optomechanical molecular cavity system. Our system leverages a spinning whispering-gallery-mode (WGM) resonator, hosting molecular vibrations, and coupled to an auxiliary optical cavity, to achieve robust and tunable nonreciprocal quantum correlations. A key feature of our system is the exploitation of the Sagnac-Fizeau effect, induced by the spinning WGM resonator, to generate nonreciprocal entanglement between the cavity mode and molecular vibrations, as well as between different collective vibrational modes. Our numerical results demonstrate that this Sagnac-Fizeau effect indeed leads to a clear directionality in entanglement, with enhanced bipartite entanglement in one direction of WGM resonator rotation and suppressed entanglement in the opposite direction, confirming the nonreciprocal nature of the generated quantum correlations.

We proposed a theoretical scheme to realize nonreciprocal bipartite entanglement in a molecular cavity optomechanical system, consisting of a spinning WGM resonator hosting molecular vibrations that is coupled to an auxiliary optical cavity. The nonreciprocal entanglement is induced by the Sagnac-Fizeau effect. The optical photons experience a red or blue frequency shift by spinning the WGM resonator in CW or CCW direction. This spinning of WGM resonator gives rise to steady-state nonreciprocal entanglement. Indeed, we find that the bipartite entanglement is enhanced in one direction and is weak in the other direction. In addition, we use the bidirectional contrast ratio to show that the nonreciprocal entanglement can be switched on ($\mathcal{C}>0$) and off ($\mathcal{C}=0$). By carefully choosing the parameters, an ideal nonreciprocal entanglement can be realized. Our results suggest that nonreciprocity can be enhanced significantly at relatively high temperatures. This proposed scheme may pave the way for the realization of new nonreciprocal devices that are robust against thermal fluctuations. Furthermore, our analysis using the bidirectional contrast ratio quantitatively confirms the switchable nature of nonreciprocal entanglement in our system. By tuning the detuning of the auxiliary cavity, we have shown that the degree of nonreciprocity can be controlled and even switched off, providing a valuable degree of control for potential quantum applications. Importantly, our investigation reveals that robust nonreciprocal entanglement, particularly vibration-vibration entanglement, can be achieved even at relatively high temperatures, highlighting the resilience of our scheme to thermal decoherence, especially when employing a larger number of molecules. This robustness to temperature, combined with the use of molecular vibrations that offer high frequencies and strong coupling, distinguishes our approach from other nonreciprocal optomechanical systems and suggests potential advantages for practical implementations.

Moreover, our findings indicate a trade-off between photon-vibration and vibration-vibration entanglement with respect to the number of molecules. Although photon-vibration entanglement may be stronger for smaller molecular ensembles, vibration-vibration entanglement becomes dominant and more robust for larger numbers of molecules, offering flexibility in tailoring entanglement properties for specific quantum tasks. The tunability of nonreciprocity via auxiliary cavity detuning and the robustness of vibration-vibration entanglement in our molecular optomechanical system are key features that differentiate our work and offer potential advantages for quantum information processing and the development of novel nonreciprocal quantum devices. Compared with existing approaches in nonreciprocal cavity optomechanics and magnomechanics, our molecular system offers a unique combination of high-frequency vibrational modes, strong optomechanical coupling, and inherent tunability, potentially paving the way for compact, robust, and versatile nonreciprocal quantum devices operating at higher frequencies and with enhanced thermal resilience. This proposed scheme, therefore, not only contributes to the fundamental understanding of nonreciprocal quantum phenomena but also holds promise for advancing quantum technologies that rely on directional quantum signal processing and robust quantum correlations. Future research directions could explore the experimental realization of this scheme, investigate the impact of realistic imperfections and noise sources, and further optimize the system parameters for specific quantum information and sensing applications.

In conclusion, our theoretical study demonstrates the feasibility of generating and controlling nonreciprocal bipartite entanglement in a molecular cavity optomechanical system.  The predicted tunability, robustness, and potential for macroscopic entanglement in this system highlight its promise as a platform for future quantum technologies based on nonreciprocal quantum phenomena.

\section*{Acknowledgment}
P.D. acknowledges the Iso-Lomso Fellowship at Stellenbosch Institute for Advanced Study (STIAS), Wallenberg Research Centre at Stellenbosch University, Stellenbosch 7600, South Africa, and The Institute for Advanced Study, Wissenschaftskolleg zu Berlin, Wallotstrasse 19, 14193 Berlin, Germany. K.S. Nisar thanks the funding from Prince Sattam bin Abdulaziz University, Saudi Arabia project number (PSAU/2024/R/1445). This research was funded by Princess Nourah bint Abdulrahman University Researchers Supporting Project number (PNURSP2025R59), Princess Nourah bint Abdulrahman University, Riyadh, Saudi Arabia. The authors are thankful to the Deanship of Graduate Studies and Scientific Research at the University of Bisha for supporting this work through the Fast-Track Research Support Program.

\textbf{Author Contributions:} K. B. E. and J.-X. P. conceptualized the work and Investigated on the methodology and simulations. A.S., P.D., A.-H. A.-A. participated in Data curation, Validation, Methodology. N. A., K.S.N., and S. G.N.E. participated in formal analysis and Supervision. All authors participated equally in the Writing, discussions, editing, and review of the manuscript.

\section*{Declaration of competing interest}
The authors declare that they have no known competing financial interests or personal relationships that could have appeared to influence the work reported in this paper.

 \section*{Data Availability}
 Data will be made available on request.
\bibliography{RefMolecular}
\end{document}